\documentclass[11pt]{article}

\IfFileExists{arxiv.sty}{\usepackage{arxiv}}{\usepackage[margin=1in]{geometry}}

\usepackage[utf8]{inputenc}
\usepackage[T1]{fontenc}
\usepackage{amsmath,amssymb,amsfonts}
\usepackage{amsthm}
\usepackage{bm}
\usepackage{graphicx}
\usepackage{steinmetz}
\usepackage{colortbl}
\usepackage{xcolor}
\usepackage{natbib}
\usepackage{url}
\usepackage{hyperref}
\usepackage{orcidlink}
\usepackage{microtype}

\graphicspath{{./images/}}

\newcommand{\E}{\mathbb{E}}

\newcommand{\diag}{\operatorname{diag}}

\theoremstyle{remark}
\newtheorem{remark}{Remark}

\title{Module-structured mixture factor models for molecular subtype discovery in transcriptomic data}

\author{
Jinran Wu\orcidlink{0000-0002-2388-3614}$^{1}$\thanks{Corresponding author: \href{mailto:jinran.wu@uq.edu.au}{jinran.wu@uq.edu.au}} \quad
Geoffrey J.~McLachlan\orcidlink{0000-0002-5921-3145}$^{1}$ \quad
Saumyadipta Pyne\orcidlink{0000-0003-3470-2345}$^{2,3}$\\[0.6em]
\small $^{1}$School of Mathematics and Physics, The University of Queensland, QLD, Australia\\
\small $^{2}$Department of Statistics and Applied Probability, University of California, Santa Barbara, CA, USA\\
\small $^{3}$Health Analytics Network, MD, USA
}
\date{}

\begin{document}
\maketitle

\begin{abstract}
High-throughput gene expression data exhibit high dimensionality, complex intergene dependence, and pronounced biological heterogeneity across samples, presenting major challenges for unsupervised clustering and disease subtype discovery. We introduce a module-structured mixture factor model that combines finite mixture modeling with low-rank latent factor representations defined at the gene-module level. By explicitly modeling gene modules in both the mean and covariance structure, the proposed framework decomposes expression variability into global gene-specific effects, cluster-specific module-level shifts, latent dependence within modules, and gene-specific residual noise. An Expectation--Conditional Maximization algorithm is applied for parameter estimation, allowing stable and scalable inference in high-dimensional transcriptomic settings. This framework enables interpretable unsupervised identification of disease-associated molecular subtypes and phenotypic heterogeneity across two autoimmune diseases using a large clinical transcriptomic dataset.
\end{abstract}


\bigskip
\section{Introduction}

A defining biological feature of gene expression data, unlike typical high-dimensional data, is its pronounced modular organization. Genes typically function in coordinated groups that share regulatory programs or biological pathways, resulting in strong within-module dependence and heterogeneous signal strengths across genes~\citep{segal2004module, wang2025stmodule}. Clustering approaches that ignore this modular structure, such as methods that rely solely on marginal gene effects or make overly simplistic covariance assumptions, are prone to numerical instability, limited interpretability, and poor reproducibility. At the other extreme, fully unstructured covariance modeling is statistically infeasible and computationally prohibitive in high-dimensional settings.

Gene modules and pathways are increasingly popular for comparison and classification of diseases~\citep{Mi2019}. Groups of diseases could be characterized not only by disease-specific modules but also by core pathways sometimes shared among them. Autoimmune diseases form a group of 80-150 complex, chronic disorders that are often debilitating and have no known cures~\citep{Martorell-Marugan2021, Shen2022}. They are commonly characterized by immune responses to self-antigens, leading to tissue damage and dysfunction in several organs. Their pathogenesis is not fully understood, and they exhibit a high degree of heterogeneity in clinical and molecular phenotypes among patients~\citep{Cheng2024}. Although these disorders damage different organs and have variable clinical outcomes, studies have found that they share many risk factors and molecular mechanisms~\citep{Li2025}.

In this study, we adopt a new module-based approach to characterize the transcriptomic signatures of two prominent autoimmune diseases: systemic lupus erythematosus (SLE) and rheumatoid arthritis (RA). SLE is a particularly heterogeneous autoimmune disease characterized by aberrant immune activation, impaired regulatory cell function, antinuclear antibody production, etc.~\citep{Cojocaru2011}. SLE phenotypes are driven by the deposition of circulating immune complexes into different organs (hence deemed systemic), triggering complement system activation and local tissue damage, e.g., lupus nephritis~\citep{Patino-Martinez2026}. At the molecular level, SLE has multiple drivers, including unchecked B cell activation, a strong interferon signature, and the massive production of pathogenic autoantibodies, e.g., anti-dsDNA.  In contrast, RA is characterized by localized, chronic joint inflammation marked by chronic inflammation of the synovium, leading to complex disease heterogeneity. Its molecular drivers are dominated by hyperactivation of the JAK-STAT signaling pathway, activation of NF-$\kappa$B, and an abundance of pro-inflammatory cytokines. With its progression, RA leads to irreversible bone tissue damage, causing persistent pain and significantly impaired joint functionality~\citep{Sharif2018}.

Molecular classification studies have refined distinct functional subtypes of SLE, including interferon-driven, neutrophil-associated, and other immune-cell-related subtypes \citep{Cheng2024}. Similarly, transcriptomic analyses of RA have identified subgroups characterized by inflammatory responses, neutrophil activation, and joint-damage-related pathways. Despite substantial differences in their underlying molecular mechanisms, notable commonalities have been observed between SLE and RA patient subtypes. For example, neutrophil activity and TGF-$\beta$ signaling, both of which play important roles in immune regulation and inflammatory processes, have been implicated across multiple autoimmune diseases. Many of these subtype discoveries have been facilitated through gene modules identified by weighted gene co-expression network analysis (WGCNA)~\citep{Langfelder2008}, highlighting the importance of modular transcriptomic organization in understanding disease heterogeneity~\citep{Cheng2024}.

From a statistical perspective, finite mixture models provide a natural framework for uncovering latent disease subtypes within heterogeneous transcriptomic populations \citep{mclachlan2005analyzing, mclachlan2019finite}. However, direct Gaussian mixture modeling becomes impractical in high-dimensional settings because of the large number of covariance parameters that must be estimated. Mixtures of factor analyzers (MFA) address this challenge by representing cluster-specific (or common) covariance structures through a low-dimensional latent factor space~\citep{mclachlan2003modelling, mclachlan2000mixtures, baek2009mixtures, mclachlan2011mixtures}, thereby achieving substantial dimension reduction while preserving important dependence patterns among genes. Due to these advantages, MFA and related latent variable mixture models have been widely used for clustering high-dimensional biological data~\citep{mclachlan2019finite}. Nevertheless, conventional MFA models typically treat genes as exchangeable variables~\citep{silkwood2024leveraging} and do not explicitly account for the modular organization inherent in biological systems. As a result, module-level dependence structures and pathway-specific sources of variation may be inadequately captured, limiting both biological interpretability and the ability to characterize disease-relevant molecular mechanisms.

To address these challenges, this study aims to develop clustering models that explicitly incorporate modular structure while remaining statistically stable and computationally scalable. In this paper, we propose a module-structured mixture factor model that integrates finite mixture modeling with low-rank latent factor representations defined at the gene-module level. The proposed framework accommodates either data-driven or externally defined gene modules and embeds them directly into the mixture model to jointly capture between-cluster heterogeneity and within-module dependence. By decomposing expression variability into global gene-specific effects, cluster-specific module-level shifts, latent module-level dependence, and gene-specific residual variability, the model achieves a principled balance between modeling flexibility, biological interpretability, and computational tractability. Unlike conventional mixtures of factor analyzers that allow cluster-specific covariance structures, the proposed model employs a common module-structured covariance representation across clusters, focusing on disease heterogeneity through interpretable module-level mean shifts.

Notably, our module-structured mixture factor modeling was able to identify, in an interpretable unsupervised manner, detailed molecular subtypes and phenotypic heterogeneity within a large clinical transcriptomic dataset across two autoimmune diseases, RA and SLE. Functional interpretation of the inferred modules was done in terms of enrichment of known molecular pathways. In addition, Human Phenotype Ontology (HPO) was used to characterize these subtypes based on the corresponding clinical phenotypes. Finally, a comparative analysis of multiple module-based methods demonstrated higher-resolution clustering
and significantly improved the performance of our model. 

The remainder of the paper is organized as follows. Section~\ref{sec:modules} describes the construction of gene modules used as model inputs. Section~\ref{msmfm} introduces the proposed module-structured mixture factor model. Parameter estimation via an ECM algorithm is presented in Section~\ref{sec:ecm}, followed by the results of empirical studies illustrating the practical utility of the proposed approach in Section~\ref{sec:empirical-setting}. Section~\ref{discussion6} concludes the paper.

\section{Gene module construction}\label{sec:modules}

Let $\bm{X}=(\bm{x}_1,\ldots,\bm{x}_n)^\top \in \mathbb{R}^{n\times p}$ denote the gene expression matrix, where $\bm{x}_i=(x_{i1},\ldots,x_{ip})^\top$ represents the gene expression profile of sample $i$, and $x_{ij}$ denotes the expression level of the $j$-th gene in sample $i$. Here, $n$ and $p$ are the number of samples and genes, respectively. Before module construction, each gene is standardized across samples to have a mean of zero and unit variance. The purpose of the module construction step is not to identify biologically optimal gene modules, but rather to obtain a stable low-resolution representation of transcriptomic structure that can be incorporated into the proposed mixture factor model.
\newline
\textit{Gene filtering.} To reduce dimensionality and remove weakly informative genes, genes are ranked according to their median absolute deviation (MAD)~\citep{tibshirani1999clustering}. For gene $j$, the MAD is defined as
$$
\operatorname{MAD}_j
=
\operatorname{median}_{1\le i\le n}
\left|
x_{ij}
-
\operatorname{median}_{1\le i\le n}(x_{ij})
\right|.
$$
The MAD provides a robust measure of marginal variability that is less sensitive to extreme observations than the sample variance. The top $p^\ast$ genes with the largest MAD values are retained for subsequent analysis.
\newline
\textit{Correlation-based clustering.} Let $\mathcal J^\ast$ denote the set of retained genes. Gene modules are constructed using hierarchical agglomerative clustering based on pairwise gene correlations. For genes $j,j'\in\mathcal J^\ast$, define the dissimilarity measure
$$
d_{jj'}
=
\sqrt{1-\operatorname{corr}(\bm{X}_{\cdot j},\bm{X}_{\cdot j'})},
$$
where $\operatorname{corr}(\cdot,\cdot)$ denotes the Pearson correlation coefficient computed across samples. Average linkage clustering is then applied to the resulting dissimilarity matrix to construct a hierarchical dendrogram.
\newline
\textit{Module definition.} The dendrogram is partitioned at a prespecified cutting threshold, yielding a collection of candidate gene modules. Modules containing fewer than $m_{\min}$ genes are removed to avoid unstable estimation arising from extremely small groups. Let $G$ denote the number of retained modules. Define
$$
c(j)\in\{1,\ldots,G\}
$$
as the module membership indicator for gene $j$, and let
$$
\mathcal G_g
=
\{j\in\mathcal J^\ast : c(j)=g\},
\qquad
g=1,\ldots,G,
$$
denote the index set corresponding to module $g$. The collection
$$
\{\mathcal G_1,\ldots,\mathcal G_G\}
$$
forms a partition of the retained gene set $\mathcal J^\ast$. For notational convenience, we write $j\in g$ whenever $j\in\mathcal G_g$.

The resulting modules are used solely to define the structured mean and covariance components of the proposed mixture factor model. Importantly, the subsequent modeling and estimation framework is agnostic to the particular module construction procedure. Consequently, alternative module definitions, including pathway-based annotations, curated biological gene sets, and externally supplied module structures, can be incorporated without modification to the model formulation or estimation algorithm.

\section{Module-structured mixture factor model}\label{msmfm}

Assume a finite mixture model with $K$ latent clusters. Let $z_i \in \{1,\ldots,K\}$ denote the cluster indicator for sample $i$, with
\[
\Pr(z_i = k)=\pi_k,
\qquad
\sum_{k=1}^K \pi_k=1.
\]
Genes are partitioned into $G$ non-overlapping modules. Let $c(j)\in\{1,\ldots,G\}$ denote the module index for gene $j$.

\subsection{Hierarchical generative model}

Conditional on cluster membership $z_i=k$, we adopt a module-structured factor-analytic model:
\begin{align}
\bm{u}_i\mid(z_i=k)
&\sim\mathcal N(\bm 0,\bm I_q),\\
\bm{x}_i\mid(\bm{u}_i,z_i=k)
&=
\bm{\delta}
+
\bm{M}\bm{\alpha}_k
+
\bm{B}\bm{u}_i
+
\bm{\varepsilon}_i,\\
\bm{\varepsilon}_i
&\sim
\mathcal N(\bm0,\bm D).
\end{align}

where $\bm{x}_i\in\mathbb R^p$ denotes the gene expression vector for sample $i$, $\bm{\delta}\in\mathbb R^p$ is a global gene-specific mean vector capturing baseline expression levels shared across all samples, and $\bm{u}_i\in\mathbb R^q$ is a $q$-dimensional latent factor. The cluster-specific effects are defined at the module level by
$\bm{\alpha}_k=(\alpha_{k1},\ldots,\alpha_{kG})^\top\in\mathbb R^G$. The residual covariance is diagonal with
$\bm D=\diag(\bm\psi)$, where
$\bm\psi=(\psi_1,\ldots,\psi_p)^\top$
collects gene-specific residual variances.

To encode the module structure, the loading matrix
$\bm B\in\mathbb R^{p\times q}$
is constrained as
\begin{equation}
\bm B=\bm S\bm M\bm H.
\end{equation}

Here
$\bm S=\diag(s_1,\ldots,s_p)$
contains the gene-specific scaling coefficients,
$\bm M\in\{0,1\}^{p\times G}$
is a gene-module assignment matrix with
$(\bm M)_{jg}=1$
if and only if
$c(j)=g$,
and
$\bm H=(\bm h_1,\ldots,\bm h_G)^\top\in\mathbb R^{G\times q}$
collects the module-level loading direction vectors.

For identifiability, we impose
$\|\bm h_g\|_2=1$
for all
$g=1,\ldots,G$. Equivalently, the model can be written gene-wise as
\[
x_{ij}
=
\delta_j
+
\alpha_{k,c(j)}
+
\bm b_j^\top\bm u_i
+
\varepsilon_{ij},
\qquad
\varepsilon_{ij}\sim\mathcal N(0,\psi_j),
\]
where
$\bm b_j$
denotes the $j$th row vector of
$\bm B$
and satisfies
\[
\bm b_j=s_j\,\bm h_{c(j)}.
\]

\subsection{Marginal cluster-conditional distribution}

Integrating out the latent factor $\bm u_i$ yields
\begin{equation}
\bm x_i\mid z_i=k
\sim
\mathcal N(\bm\mu_k,\bm\Sigma),
\end{equation}
with
\begin{equation}
\bm\mu_k
=
\bm\delta+\bm M\bm\alpha_k,
\qquad
\bm\Sigma
=
\bm B\bm B^\top+\bm D.
\end{equation}
\newline
\begin{remark}\label{remark1}
The decomposition $\bm{\mu}_k=\bm{\delta}+\bm{M}\bm{\alpha}_k$ is not unique, since adding a common constant to the baseline expression of a module and subtracting the same constant from the corresponding module effects leaves the marginal mean unchanged. To ensure identifiability, we impose the module-wise constraint $\sum_{k=1}^{K}\pi_k\alpha_{kg}=0, g=1,\ldots,G,$ where $\pi_k$ denotes the mixing proportion of cluster $k$. Under this constraint, $\bm{\delta}$ represents the global baseline gene expression, while $\alpha_{kg}$ represents the cluster-specific deviation of module $g$
from this baseline.
\end{remark}

Under
$\bm B=\bm S\bm M\bm H$
and
$\bm D=\diag(\bm\psi)$,
$\bm\Sigma$ becomes
\begin{equation}
\bm\Sigma
=
\diag(\bm\psi)
+
\bm S\bm M\bm H\bm H^\top\bm M^\top\bm S.
\end{equation}

\subsection{Observed-data mixture likelihood}

The marginal density of
$\bm x_i$
is a finite mixture of $K$ Gaussian distributions,
\begin{equation}
p(\bm x_i\mid\bm\Theta)
=
\sum_{k=1}^K
\pi_k
\phi_p(\bm x_i;\bm\mu_k,\bm\Sigma),
\end{equation}
where
$\bm\Theta$
denotes the full collection of model parameters.

The observed-data log-likelihood is therefore
\begin{equation}
\ell(\bm\Theta)
=
\sum_{i=1}^n
\log
\left[
\sum_{k=1}^K
\pi_k
\phi_p(\bm x_i;\bm\mu_k,\bm\Sigma)
\right].
\end{equation}

\subsection{Interpretation}

The proposed model admits a clear interpretation as a module-structured mixture of factor analyzers, in which gene expression variability is decomposed into four distinct and interpretable components.

\begin{enumerate}

\item Global gene-specific baseline effects ($\bm\delta$), which capture systematic expression differences shared across all samples and clusters. These effects account for gene-level baseline activity and remove global expression offsets before clustering and covariance modeling.

\item Cluster-specific module shifts ($\bm\alpha_k$), which represent differential activation or suppression of entire gene modules across latent clusters. By operating at the module level, these mean shifts encode biologically meaningful between-cluster heterogeneity at the level of pathways or functional groups, rather than individual genes, thereby improving both interpretability and statistical efficiency.

\item Latent-factor dependence within modules ($\bm B,\bm u_i$), which models residual correlation among genes through a low-rank factor-analytic structure. The loading matrix $\bm B=\bm S\bm M\bm H$ imposes structured parameter sharing, so that genes within the same module share common loading directions while differing in magnitude through gene-specific scaling coefficients. This component captures coordinated expression patterns beyond mean shifts, such as shared regulatory programs or unobserved cellular states, while substantially reducing the effective dimensionality of the covariance structure.

\item Gene-specific residual variability ($\bm\psi$), which accounts for idiosyncratic noise and measurement error not explained by the latent factors or module structure. Allowing heterogeneous residual variances across genes avoids overly restrictive homoscedastic assumptions and improves model flexibility.

\end{enumerate}

By jointly modeling module-level structure in both the mean and covariance, the proposed framework achieves substantial dimension reduction while retaining the expressive power of a mixture of factor analyzers. The explicit separation of global effects, cluster-specific mean structure, latent dependence, and gene-specific noise enhances interpretability, improves estimation stability in high-dimensional settings, and facilitates biologically meaningful clustering of transcriptomic profiles.

\begin{remark}
The proposed parameterization achieves a substantial reduction in the number of covariance-related parameters. Whereas a conventional mixture of factor analyzers requires estimation of a loading matrix with $pq$ free loading parameters, the module-structured representation $\bm B=\bm S\bm M\bm H$ reduces the effective number of loading parameters to approximately $p+Gq$. Thus, when $G\ll p$, the complexity of the covariance model is substantially reduced. This parsimonious representation is particularly advantageous in high-dimensional transcriptomic applications, where the number of genes is often considerably larger than the number of biologically meaningful modules. This reduction in parameter complexity improves estimation stability and computational scalability while preserving biologically interpretable dependence structures.
\end{remark}

\section{Estimation via an ECM algorithm}\label{sec:ecm}

Parameter estimation is carried out using an ECM algorithm~\citep{meng1993maximum, mclachlan2008algorithm, ng2011algorithm}. The complete data consist of the observed vectors together with the latent cluster indicators $\{z_i\}_{i=1}^n$ and the latent factors $\{\bm{u}_i\}_{i=1}^n$. The ECM framework allows the maximization step to be decomposed into a sequence of lower-dimensional conditional maximization problems, leading to improved numerical stability in high dimensions.

Let
\[
\bm{\Theta}
=
(\bm{\pi}, \bm{\delta}, \bm{\alpha},
\bm{H}, \bm{s}, \bm{\psi})
\]
denote the full parameter set. Recall the module-structured factor-analytic representation
\[
\bm{x}_i
=
\bm{\mu}_{z_i}
+
\bm{B}\bm{u}_i
+
\bm{\varepsilon}_i,
\qquad
\bm{\varepsilon}_i\sim\mathcal{N}(\bm{0},\bm{D}),
\]
where
\[
\bm{\mu}_k=\bm{\delta}+\bm{M}\bm{\alpha}_k,\qquad
\bm{B}=\bm{S}\bm{M}\bm{H},\qquad
\bm{D}=\diag(\bm{\psi}).
\]

\subsection{E-step}

Given current parameter estimates $\bm{\Theta}^{(t)}$, the E-step computes conditional expectations with respect to the posterior distribution of the missing data.

\textit{Posterior cluster probabilities.}  The posterior probability that observation $i$ belongs to cluster $k$ is
\begin{equation}
r_{ik}
=
\Pr(z_i = k \mid \bm{x}_i; \bm{\Theta}^{(t)})
=
\frac{
\pi_k^{(t)}
\,
\phi_p\!\left(
\bm{x}_i;
\bm{\mu}_k^{(t)},
\bm{\Sigma}^{(t)}
\right)
}{
\sum_{\ell=1}^K
\pi_\ell^{(t)}
\,
\phi_p\!\left(
\bm{x}_i;
\bm{\mu}_\ell^{(t)},
\bm{\Sigma}^{(t)}
\right)
}.
\end{equation}
where
\[
\bm{\mu}_k^{(t)} = \bm{\delta}^{(t)} + \bm{M}\bm{\alpha}_k^{(t)},
\qquad
\bm{\Sigma}^{(t)} = \bm{B}^{(t)}(\bm{B}^{(t)})^\top + \bm{D}^{(t)},
\]
with $\bm{B}^{(t)}=\bm{S}^{(t)}\bm{M}\bm{H}^{(t)}$ and
$\bm{D}^{(t)}=\diag(\bm{\psi}^{(t)})$.
\newline

\textit{Conditional moments of latent factors.} Conditional on $z_i = k$, the posterior distribution of the latent factor $\bm{u}_i$ is Gaussian,
\begin{equation}
\bm{u}_i \mid (\bm{x}_i, z_i=k)
\sim
\mathcal{N}\!\left(
\bm{m}_{ik}, \bm{V}
\right),
\end{equation}
with the standard factor-analysis expressions
\begin{equation}
\bm{V}
=
\left(
\bm{I}_q + \bm{B}^\top \bm{D}^{-1}\bm{B}
\right)^{-1},
\qquad
\bm{m}_{ik}
=
\bm{V}\bm{B}^\top\bm{D}^{-1}
\left(\bm{x}_i-\bm{\mu}_k\right).
\label{eq:posterior_u}
\end{equation}
For later use, define
\begin{equation}
\bm{C}_{ik}
=
\E(\bm{u}_i \bm{u}_i^\top \mid \bm{x}_i, z_i=k)
=
\bm{V} + \bm{m}_{ik} \bm{m}_{ik}^\top .
\label{eq:Cik}
\end{equation}
\newline

\textit{Structured computation under the module parameterization.} Let $\bm{e}_{ik}=\bm{x}_i-\bm{\mu}_k$ and denote $e_{ikj}$ its $j$th element.
Using $\bm{B}=\bm{S}\bm{M}\bm{H}$ and $\bm{D}=\diag(\bm{\psi})$, we have
\begin{equation}
\bm{B}^\top\bm{D}^{-1}\bm{B}
=
\sum_{g=1}^G
\left(
\sum_{j \in g}
\frac{s_j^2}{\psi_j}
\right)
\bm{h}_g \bm{h}_g^\top
\;\equiv\;
\sum_{g=1}^G w_g\,\bm{h}_g\bm{h}_g^\top,
\qquad
w_g=\sum_{j\in g}\frac{s_j^2}{\psi_j}.
\label{eq:BtDB}
\end{equation}
and
\begin{equation}
\bm{T}_{ik}
\;\equiv\;
\bm{B}^\top\bm{D}^{-1}\bm{e}_{ik}
=
\sum_{g=1}^G
\left(
\sum_{j \in g}
\frac{s_j}{\psi_j} e_{ikj}
\right)
\bm{h}_g
\;\equiv\;
\sum_{g=1}^G c_{ikg}\,\bm{h}_g,
\qquad
c_{ikg}=\sum_{j\in g}\frac{s_j}{\psi_j}e_{ikj}.
\label{eq:Tik}
\end{equation}
Then $\bm{m}_{ik}=\bm{V}\bm{T}_{ik}$, with
\[
\bm{V}
=
\left(
\bm{I}_q+\sum_g w_g\bm{h}_g\bm{h}_g^\top
\right)^{-1}.
\]

\subsection{CM-steps}

Given the conditional expectations from the E-step, the M-step is carried out via a sequence of conditional maximization steps.

\textit{CM-step 1: Mixing proportions.}
\begin{equation}
\pi_k^{(t+1)}
=
\frac{1}{n}
\sum_{i=1}^n r_{ik}.
\end{equation}
\newline

\textit{CM-step 2: Cluster-specific module effects.} Fix $(\bm{\delta},\bm{S},\bm{H},\bm{\psi})$.
For module $g$, define $a_{ikg}=\bm{h}_g^\top\bm{m}_{ik}$.
The first-order condition for $\alpha_{kg}$ yields the closed-form update
\begin{equation}
\alpha_{kg}^{(t+1)}
=
\frac{
\sum_{i=1}^n r_{ik}\sum_{j\in g}\psi_j^{-1}
\left(
x_{ij}-\delta_j^{(t)}-s_j^{(t)} a_{ikg}
\right)
}{
\sum_{i=1}^n r_{ik}\sum_{j\in g}\psi_j^{-1}
}.
\label{eq:alpha-update}
\end{equation}
After updating $\alpha_{kg}$, the module effects are centered using $\pi_k^{(t+1)}$ to satisfy the identifiability constraint in Remark~\ref{remark1}.
\newline

\textit{CM-step 3: Global gene means.} Fix $(\bm{\alpha},\bm{S},\bm{H},\bm{\psi})$.
For gene $j$ in module $g=c(j)$, with $a_{ikg}=\bm{h}_g^\top\bm{m}_{ik}$,
\begin{equation}
\delta_j^{(t+1)}
=
\frac{1}{n}
\sum_{i=1}^n
\left[
x_{ij}
-
\sum_{k=1}^K r_{ik}
\left(
\alpha_{k,g}^{(t+1)}
+
s_j^{(t)} a_{ikg}
\right)
\right].
\label{eq:delta-update}
\end{equation}
\newline

\textit{CM-step 4: Update of $s_j$ (gene-specific scaling).} Fix $(\bm{\delta},\{\bm{\alpha}_k\}_{k=1}^K,\{\bm{h}_g\}_{g=1}^G,\bm{\psi})$.
For gene $j$ in module $g=c(j)$, define
\[
e_{ikj}=x_{ij}-\delta_j-\alpha_{k,g},
\qquad
\bm{m}_{ik}=\mathbb E(\bm{u}_i\mid \bm{x}_i,z_i=k),
\qquad
\bm{C}_{ik}
=
\mathbb E(\bm{u}_i\bm{u}_i^\top\mid \bm{x}_i,z_i=k)
=
\bm{V}
+
\bm{m}_{ik}\bm{m}_{ik}^\top,
\]
and
\[
a_{ikg}
=
\bm{h}_g^\top \bm{m}_{ik},
\qquad
b_{ikg}
=
\bm{h}_g^\top \bm{C}_{ik}\bm{h}_g
=
\bm{h}_g^\top \bm{V}\bm{h}_g
+
(\bm{h}_g^\top\bm{m}_{ik})^2.
\]
The part of the $Q$-function involving $s_j$ is quadratic, and the first-order condition yields
\[
s_j^{(t+1)}
=
\frac{\sum_{i=1}^n\sum_{k=1}^K r_{ik}\,e_{ikj}\,a_{ikg}}
{\sum_{i=1}^n\sum_{k=1}^K r_{ik}\,b_{ikg}},
\qquad
g=c(j).
\]
Since the representation $\bm{b}_j=s_j\bm{h}_{c(j)}$ is invariant to a common sign change within each module, a module-level sign convention is imposed on $\bm{h}_g$ for identifiability.
\newline

\textit{CM-step 5: Update of $\bm{h}_g$ (module loading direction).}
Fix $(\bm{\delta},\{\bm{\alpha}_k\}_{k=1}^K,\bm{S},\bm{\psi})$.
For module $g$, retaining only the terms of the $Q$-function that depend on $\bm{h}_g$, the corresponding optimization problem is
\[
\max_{\|\bm{h}_g\|_2=1}
\left\{
-\frac{1}{2}\bm{h}_g^\top \bm{A}_g \bm{h}_g
+
\bm{b}_g^\top \bm{h}_g
\right\}.
\]
Let
\[
w_g=\sum_{j\in g}\frac{s_j^2}{\psi_j},
\qquad
c_{ikg}=\sum_{j\in g}\frac{s_j}{\psi_j}\,e_{ikj}.
\]
Then $\bm{A}_g$ and $\bm{b}_g$ admit the explicit expressions
\[
\bm{A}_g
=
\sum_{i=1}^n\sum_{k=1}^K r_{ik}\, w_g\, \bm{C}_{ik},
\qquad
\bm{b}_g
=
\sum_{i=1}^n\sum_{k=1}^K r_{ik}\, c_{ikg}\, \bm{m}_{ik}.
\]
Therefore,
\[
\bm{h}_g^{(t+1)}
=
\arg\max_{\|\bm{h}\|_2=1}
\Big\{
-\tfrac12 \bm{h}^\top \bm{A}_g \bm{h}
+
\bm{b}_g^\top \bm{h}
\Big\}.
\]
The KKT conditions imply that there exists $\lambda\in\mathbb R$ such that
\[
(\bm{A}_g+\lambda \bm{I}_q)\,\bm{h}_g=\bm{b}_g,
\qquad
\|\bm{h}_g\|_2=1,
\]
hence
\[
\bm{h}_g(\lambda)
=
(\bm{A}_g+\lambda \bm{I}_q)^{-1}\bm{b}_g
\]
and $\lambda$ solves
\[
\phi(\lambda)
=
\|\bm{h}_g(\lambda)\|_2^2-1
=
0,
\qquad
\lambda>-\lambda_{\min}(\bm{A}_g).
\]
In practice, one may compute $\lambda$ by a one-dimensional root-finding method (e.g., bisection),
and then set
\[
\bm{h}_g^{(t+1)}
=
\bm{h}_g(\lambda).
\]
Afterwards, enforce the sign convention $h_{g,m(g)}>0$ where
$m(g)=\arg\max_\ell|h_{g\ell}|$.
\newline

\textit{CM-step 6: Residual variances.}
Let $g=c(j)$ and recall
$e_{ikj}=x_{ij}-\delta_j-\alpha_{k,g}$ and
$a_{ikg}=\bm{h}_g^\top\bm{m}_{ik}$,
where $\bm{m}_{ik}=\mathbb E(\bm{u}_i\mid \bm{x}_i,z_i=k)$ and
$\bm{V}=\mathrm{cov}(\bm{u}_i\mid \bm{x}_i,z_i=k)$ from the E-step.
Using $\bm{C}_{ik}=\bm{V}+\bm{m}_{ik}\bm{m}_{ik}^\top$, we have
\[
\bm{h}_g^\top\bm{C}_{ik}\bm{h}_g
=
\bm{h}_g^\top\bm{V}\bm{h}_g
+
a_{ikg}^2.
\]
Then the gene-specific residual variances are updated by
\begin{equation}
\psi_j^{(t+1)}
=
\frac{1}{n}
\sum_{i=1}^n
\sum_{k=1}^K
r_{ik}
\left[
e_{ikj}^2
-
2 s_j^{(t+1)} e_{ikj} a_{ikg}
+
(s_j^{(t+1)})^2
\left(
\bm{h}_g^\top\bm{V}\bm{h}_g
+
a_{ikg}^2
\right)
\right],
\label{eq:psi-update}
\end{equation}
where we use $\sum_{k=1}^K r_{ik}=1$ for each $i$. In practice, we bound $\psi_j^{(t+1)}$ below by a small positive constant to prevent degeneracy.\\

\textit{Initialization.} The ECM algorithm is initialized using a reduced-dimension representation based on gene modules. Specifically, for each sample, we compute module-level mean expression profiles and apply $K$-means clustering (with multiple random starts) in the resulting $G$-dimensional space to obtain an initial hard partition. The initial responsibilities are set to $r_{ik}^{(0)}=\mathbb{I}\{\hat z_i^{(0)}=k\}$, and the mixing proportions are initialized accordingly. Given the initial cluster assignments, the global gene means are initialized by sample averages, $\delta_j^{(0)}=n^{-1}\sum_{i=1}^n x_{ij}$, and the cluster-specific module effects are initialized as within-cluster averages of the module-level centered expression profiles. The module loading directions $\{\bm{h}_g\}_{g=1}^G$ are initialized as independent random unit vectors in $\mathbb{R}^q$, and the gene-specific scaling coefficients are set to $s_j^{(0)}=1$. Finally, the residual variances are initialized using gene-wise mean squared residuals under the initial mean structure, with a small positive lower bound imposed to avoid degeneracy.

\section{Results}
\label{sec:empirical-setting}

\subsection{Autoimmune disease data}

The empirical analyses are based on a large transcriptomic dataset comprising samples drawn from two clinically distinct autoimmune diseases, RA and SLE. The anonymized publicly available dataset (entry GSE45291) was obtained from the ADEx database (Autoimmune Diseases Explorer~\citep{Martorell-Marugan2021}). It contains a total of 785 samples, including 493 RA samples and 292 SLE samples, with raw expression measurements available for 9,671 genes. While the disease membership of each sample is known a priori, this information is not used at any stage of model fitting, model selection, or subtype inference. Instead, all analyses are conducted in an unsupervised manner based solely on the observed gene expression profiles, allowing latent molecular structure and heterogeneity to emerge from the data without imposing disease-level labels.

To reduce dimensionality and improve estimation stability in high-dimensional transcriptomic data, genes are first screened using the MAD, and only highly variable genes are retained for downstream analysis. Gene expression values are subsequently standardized to have zero mean and unit variance to ensure comparability across genes and samples. The retained genes are then grouped into co-expression modules using hierarchical clustering based on pairwise gene-gene correlations, with average-linkage clustering applied to a correlation-based dissimilarity matrix. Small or weakly defined clusters are removed to enhance the robustness and interpretability of module-level estimation.

After preprocessing under this fixed empirical protocol, the final analysis set consists of 979 highly variable genes organized into 19 co-expression modules. These gene modules are constructed as a preprocessing step and treated as fixed inputs to the proposed model. The resulting processed dataset serves as the input for all subsequent model-based analyses, including subtype discovery, module-level effect estimation, and functional interpretation. Detailed implementation choices are provided in Appendix~\ref{app:module_settings}.

\subsection{Model selection}

\begin{figure}[ht]
\centering
\includegraphics[width=0.55\textwidth]{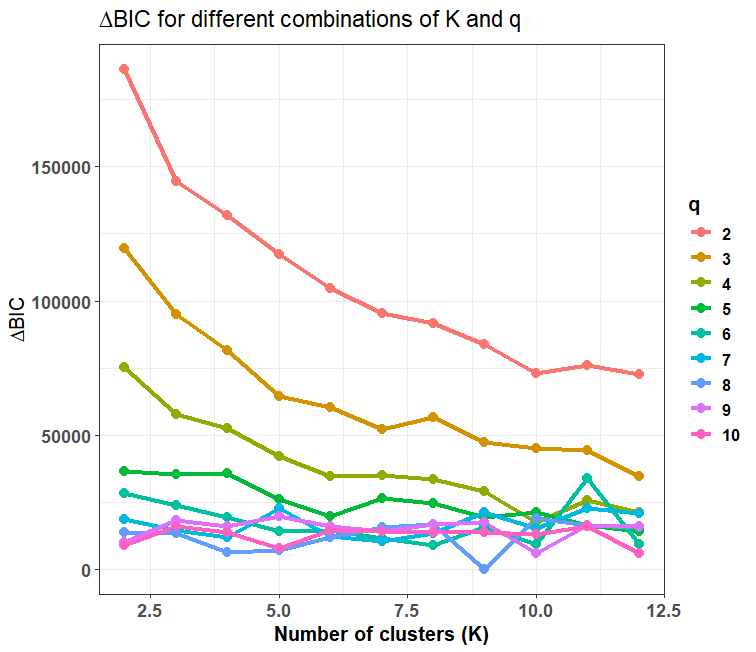}
\caption{$\Delta\text{BIC}$ values for the dataset across different numbers of mixture components $K$ and latent factor dimensions $q$.}
\label{fig:BIC}
\end{figure}

The proposed module-structured mixture factor model is fitted to the combined dataset over a grid of candidate values for the number of mixture components $K$ and the latent factor dimension $q$. Model selection is carried out using the Bayesian information criterion (BIC), with smaller values indicating a better trade-off between model fit and complexity. Figure~\ref{fig:BIC} displays the resulting $\Delta$BIC values across the $(K,q)$ grid, where $\Delta$BIC is defined as the difference between a model's BIC and the minimum BIC among all candidate models. Consequently, the model with the smallest BIC has $\Delta$BIC = 0. 

The optimal model configuration is selected as the combination of $(K,q)$ that minimizes the BIC and is used for all subsequent analyses. For the combined RA-SLE dataset, the minimum BIC is attained at $K = 9$ and $q = 8$.

\subsection{Clustering structure and disease-associated heterogeneity}

Using the selected model configuration $(K = 9, q = 8)$, samples from the dataset are assigned to latent clusters according to posterior cluster responsibilities. Table~\ref{tab:RA-SLE-clusters-transposed} summarizes the distribution of samples across the inferred clusters, stratified by disease status. Although disease labels are not used during model fitting or clustering, the inferred clusters exhibit a pronounced disease-specific composition. Each cluster is dominated by samples from a single disease, with clusters 1, 5, 6, 7, and 9 primarily composed of RA samples, and clusters 2, 3, 4, and 8 almost exclusively populated by SLE samples. This complete separation indicates that disease-level molecular structure~\citep{Cheng2024} emerges naturally from the unsupervised analysis, while the presence of multiple clusters within each disease highlights substantial within-disease heterogeneity.

\begin{table}[ht]
\centering
\small
\caption{Distribution of samples across inferred clusters by disease status.}
\label{tab:RA-SLE-clusters-transposed}
\begin{tabular}{c c c c c c c c c c}
\hline
 & \multicolumn{9}{c}{Cluster} \\
Disease 
& 1 & 2 & 3 & 4 & 5 & 6 & 7 & 8 & 9 \\
\hline
RA  & 15 & 0 & 0 & 0 & 149 & 66 & 228 & 0 & 35 \\
SLE & 0 & 9 & 114 &113 & 0 & 0 & 0 & 56 & 0 \\
\hline
\end{tabular}
\end{table}

\begin{figure}[!ht]
\centering
\includegraphics[width=\textwidth]{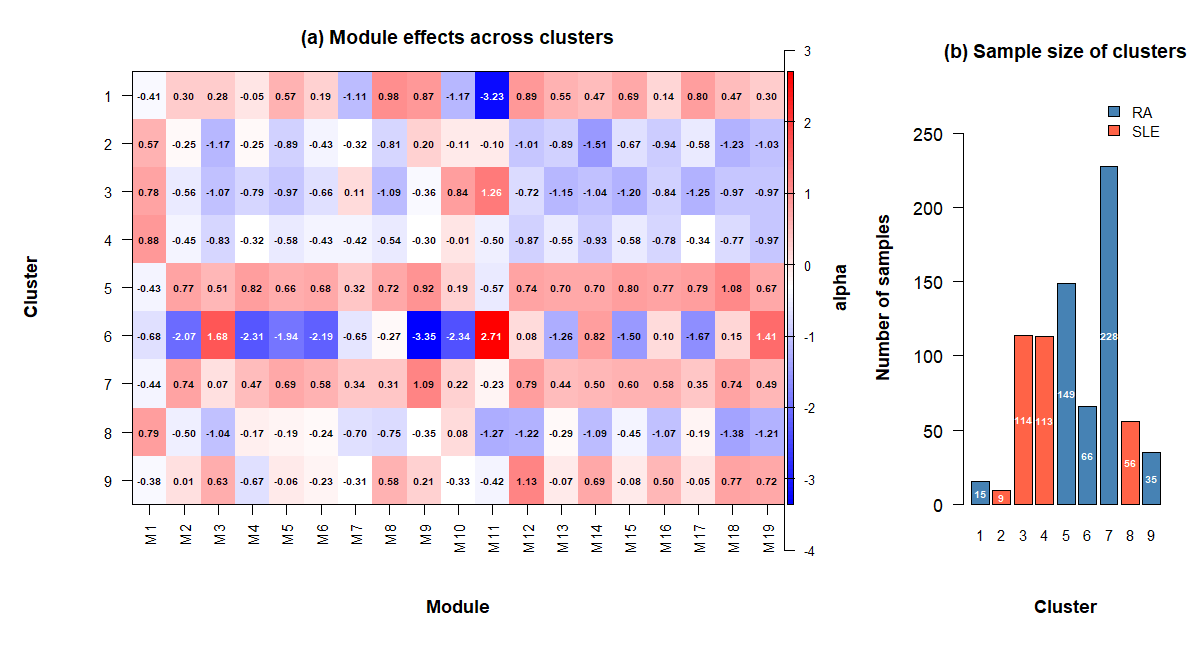}
\caption{
Cluster-specific module effects and cluster prevalence in the investigated dataset.
(a) Heatmap of estimated module-level mean effects $\alpha_{k,g}$, with rows corresponding to inferred clusters and columns corresponding to gene modules. Numerical values indicate estimated effects rounded to two decimal places.
(b) Number of samples assigned to each inferred cluster.
}
\label{fig:cluster-alpha}
\end{figure}

Figure~\ref{fig:cluster-alpha} further characterizes the inferred clusters in terms of their module-level expression patterns and sample sizes. Panel~(a) displays the estimated cluster-specific module effects $\alpha_{k,g}$, where rows correspond to inferred clusters and columns correspond to gene modules. Clear and structured variation is observed across clusters, with many clusters exhibiting coordinated up- or down-regulation across multiple modules rather than isolated deviations in individual modules.

Several clusters exhibited clearly distinct module-level effect patterns, which can be summarized in terms of characteristic signature modules. Clusters~2, 3, 4, and~8 (SLE-dominated) were characterized by broadly negative signatures, particularly across Modules~3-6 and Modules~10-15, where effect sizes frequently fell below $-1.0$ (e.g., Cluster~2: Module~14 $\alpha = -1.51$; Cluster~3: Modules~12-14 $\alpha = -1.15, -1.04$). In contrast, RA-dominated clusters such as Clusters~5 and~7 showed positive signature modules spanning Modules~5, 8, 9, and~15-18, with effect sizes typically between $0.6$ and $1.1$ (e.g., Cluster~5: Module~9 $\alpha = 0.92$, Module~18 $\alpha = 1.08$; Cluster~7: Module~9 $\alpha = 1.09$). These patterns indicate a clear separation between globally suppressive (SLE) and globally activated (RA) module signatures.

Within RA clusters, additional heterogeneity was observed in the composition of signature modules. Cluster~6 displayed a distinctive bidirectional signature, with strongly negative modules (Modules~2, 4, 8-10; e.g., Module~9 $\alpha = -3.35$) co-occurring with strongly positive modules (Modules~3 and~11; $\alpha = 1.68$ and $2.71$, respectively), suggesting a dysregulated or reprogrammed molecular state. Cluster~5, by contrast, showed a coherent positive signature across nearly all modules, whereas Cluster~1 and Cluster~9 exhibited mixed but weaker signatures, with moderate positive effects in Modules~8-9 ($\alpha= 0.98, 0.87$ in Cluster~1) alongside negative values in Modules~10-11. Among SLE clusters, Cluster~3 showed partial deviation from the global negative pattern, with a local positive signature in Module~11 ($\alpha = 1.26$), indicating that even within SLE, there exist substructures defined by specific module activation.

Despite the overall separation between RA and SLE clusters, limited overlap in signature modules was observed. For example, moderate positive effects in Modules~8-9 appeared in both RA Cluster~1 and, to a lesser extent, SLE Cluster~3, suggesting shared molecular components. However, the dominant distinction lies in the directionality and magnitude of the signatures: RA clusters are primarily defined by coordinated positive module effects, whereas SLE clusters are characterized by consistent negative regulation across the same module groups. The extreme range observed in Cluster~6 ($\alpha$ from approximately $-3.35$ to $2.71$) further highlights that the principal source of heterogeneity arises from the strength and polarity of module-specific effects rather than from isolated module differences.

Panel~(b) summarizes the distribution of samples across clusters by disease status. The cluster sizes are moderately imbalanced, with a few large clusters (e.g., Cluster~7 with $n=228$ and Cluster~5 with $n=149$, both RA-dominated) and several smaller clusters (e.g., Cluster~2 with $n=9$ and Cluster~1 with $n=15$). Importantly, the distribution is strongly stratified by disease: RA samples are concentrated in Clusters~1, 5, 6, 7, and~9, whereas SLE samples are almost exclusively assigned to Clusters~2, 3, 4, and~8. This sharp partition indicates that the inferred clustering structure captures disease-associated variation while still allowing for multiple subtypes within each disease.

To sum up, the results demonstrate that the proposed model captures both global disease separation and finer-grained molecular heterogeneity. The identification of distinct signature modules across clusters, combined with the non-trivial and disease-aligned cluster sizes, suggests that the inferred subgroups reflect biologically meaningful molecular programs rather than artefacts of sample distribution. These findings highlight the presence of multiple, functionally distinct molecular subtypes within RA and SLE, characterized by differences in both the direction and magnitude of module-level effects.

\subsection{Functional interpretation of inferred modules}

\subsubsection{Functional annotation and disease-associated cluster-level interpretation}

To characterize the biological processes underlying the inferred clusters, Gene Ontology (GO) and KEGG pathway enrichment analyses were performed for each gene module and integrated with the cluster-specific effects $\alpha_{k,g}$. The results are presented in Table~\ref{tab:module-functional-summary}.

A clear separation between RA- and SLE-associated molecular programs is observed at the module level. SLE-dominant clusters (Clusters~2, 3, 4, and~8) exhibit broadly negative $\alpha$ values across multiple modules, particularly those enriched for immune signaling and inflammatory pathways, including Modules~2, 9, 15, and~17. In these clusters, effect sizes frequently fall below $-1.0$, indicating coordinated suppression of immune and signaling processes. In contrast, RA-associated clusters (Clusters~1, 5, 6, 7, and~9) display more heterogeneous but generally positive module-level effects. Clusters~5 and~7 show coordinated up-regulation across modules related to ribosome biogenesis, RNA processing, and protein modification (e.g., Modules~4, 5, and~12), with $\alpha$ values typically ranging from $0.5$ to $1.1$, suggesting enhanced biosynthetic and regulatory activity. Other RA clusters (e.g., Clusters~1 and~9) exhibit more moderate and mixed patterns, indicating intermediate molecular states.

Beyond disease-level separation, the $\alpha$ patterns reveal substantial heterogeneity across subtypes. Among RA clusters, Cluster~6 is particularly distinctive, showing a strongly bidirectional signature with pronounced negative effects in some modules (e.g., Module~9: $\alpha = -3.35$) and strong positive effects in others (e.g., Module~11: $\alpha = 2.71$), consistent with a highly dysregulated phenotype. In comparison, Cluster~5 shows more uniform activation, whereas Cluster~7 exhibits moderate but consistent up-regulation. SLE-associated clusters also display heterogeneity. While Clusters~2 and~4 show consistent suppression across most modules, Cluster~3 exhibits localized deviations, including moderate positive effects in specific modules, and Cluster~8 displays a comparatively attenuated pattern. These results indicate that SLE subtypes are primarily distinguished by differences in the magnitude and localisation of module suppression.

Although limited overlap in module-level activity is observed between diseases, occasional similarities exist (e.g., moderate activity in Modules~8-9 across selected RA and SLE clusters). However, the dominant distinction lies in directionality: RA clusters are characterized by coordinated activation, whereas SLE clusters exhibit systematic suppression of similar functional modules. Together, these results demonstrate that both disease-level differences and subtype heterogeneity arise from structured shifts across gene modules. From a clinical perspective, the identification of distinct module-driven subtypes suggests potential strategies for patient stratification and targeted intervention.

\begin{table}[!ht]
\centering
\caption{Functional annotation summary of inferred gene modules.}
\label{tab:module-functional-summary}
\begin{tabular}{c c c p{8cm}}
\hline
Module & Size & Mapped genes & Representative functional themes (GO / KEGG) \\
\hline
1  & 250 & 243 & Cell adhesion regulation; hypoxia response; PI3K-Akt, FoxO and mTOR signalling; Th17 differentiation \\
2  & 105 & 100 & Myeloid activation; ROS metabolism; exocytosis; chemokine signaling; innate immune response \\
3  & 70  & 67  & Telomere maintenance; chromosome organization; oxidative phosphorylation; apoptosis \\
4  & 28  & 28  & Ribosome biogenesis; translation; rRNA processing; spliceosome \\
5  & 23  & 23  & RNA transport; nuclear export; nucleocytoplasmic transport; spliceosome \\
6  & 50  & 47  & Viral processes; translation; mitochondrial gene expression; protein folding \\
7  & 22  & 22  & Ion homeostasis; calcium signaling; stress response; chromatin organization \\
8  & 66  & 57  & Nervous and vascular development; membrane potential regulation; signaling pathways \\
9  & 27  & 25  & IL-6 production; NF-$\kappa$B signaling; apoptosis; inflammatory response \\
10 & 20  & 20  & Muscle contraction; platelet activation; coagulation; vascular processes \\
11 & 69  & 65  & Catabolic regulation; RNA stability; Wnt signaling; metabolic pathways \\
12 & 53  & 52  & Protein ubiquitination; post-translational modification; vesicle organization \\
13 & 34  & 33  & Nucleotide biosynthesis; purine metabolism; protein folding \\
14 & 24  & 24  & RNA processing; splicing; cell division; organelle fission \\
15 & 24  & 22  & T cell activation; Th1/Th2/Th17 differentiation; MAPK signaling \\
16 & 22  & 21  & Transcriptional regulation; RNA polymerase II activity; cognition-related processes \\
17 & 35  & 34  & TCR signaling; immune activation; cytokine signaling; PI3K-Akt pathway \\
18 & 33  & 33  & Intracellular transport; lysosome; apoptosis; JAK-STAT signaling \\
19 & 24  & 23  & Transcriptional repression; angiogenesis; differentiation; Rap1 signaling \\
\hline
\end{tabular}
\end{table}

\subsubsection{HPO-based phenotype enrichment analysis}

To further assess the clinical relevance of the inferred gene modules, enrichment analysis was performed using the HPO, which links genes to curated phenotype annotations~\citep{Kohler2021}. Because HPO encompasses a broad spectrum of clinical conditions, including developmental disorders and rare genetic diseases, only enrichments with potential relevance to autoimmune disease were considered for biological interpretation.

Several modules exhibited enrichment for immune- and haematological-related phenotypes (Table~\ref{tab:hpo_summary}). In particular, Modules~15 and~17 were associated with autoimmune thrombocytopenia, persistent Epstein-Barr virus (EBV) viremia, lymphadenopathy, and abnormal T-cell physiology, suggesting involvement in adaptive immune dysregulation and immune activation. These findings are broadly consistent with established pathogenic mechanisms in RA and SLE. In addition, Module~9 was enriched for infection-related phenotypes, including candidiasis and splenomegaly, whereas Module~10 was associated with thrombocytopenia and other platelet-related abnormalities, reflecting haematological manifestations frequently observed in autoimmune disease. Several additional modules were associated with developmental, neurological, or structural phenotypes. Given the broad scope of HPO annotations, these enrichments were interpreted as non-specific background signals rather than direct evidence of autoimmune disease mechanisms.

To sum up, the HPO analysis provides complementary support for the biological relevance of the inferred module structure. As summarized in Table~\ref{tab:hpo_summary}, the strongest phenotype enrichments were concentrated in immune- and haematological-related modules, reinforcing the interpretation that the identified molecular subtypes are associated with distinct immunological processes and clinically relevant disease manifestations.

\begin{table}[!ht]
\centering
\caption{Representative HPO enrichments for modules with potential relevance to autoimmune disease.}
\label{tab:hpo_summary}
\begin{tabular}{c l p{8cm}}
\hline
Module & Category & Representative HPO phenotype terms \\
\hline
9  & Infection/immune &
Candidiasis; splenomegaly; aphthous stomatitis \\
10 & Haematological &
Thrombocytopenia; platelet anisocytosis; bleeding tendency \\
15 & Immune/autoimmune &
Autoimmune thrombocytopenia; EBV viremia; splenomegaly \\
17 & Immune /T-cell &
Lymphadenopathy; abnormal T-cell physiology; recurrent pneumonia; EBV viremia \\
\hline
\end{tabular}
\end{table}

\subsection{Comparative analysis of module-based methods}

The WGCNA-based module construction was performed using a signed co-expression network. The soft-thresholding power was selected based on the scale-free topology criterion, and when no clear optimum was identified, a default value of $\beta = 6$ was adopted. The minimum module size was set to 20 genes, and the module merging threshold was set to 0.25. Pearson correlation was used to construct the network. To ensure numerical stability and compatibility with downstream modeling, module reassignment and kME-based filtering were disabled. Genes assigned to the grey module were excluded from further analysis.

To evaluate the impact of module construction on downstream clustering, we compared the MAD-HC and WGCNA approaches using the same set of pre-selected genes. The MAD-HC method identified 19 gene modules from 979 genes, whereas WGCNA resulted in only 3 modules after excluding 4 genes assigned to the grey module. These modules were then used as input to the same MSD-MFM-ECM framework. Model selection was performed via a grid search over $K$ (number of clusters) and $q$ (latent dimension), using our proposed method.

The model selection results are summarized in Table~\ref{tab:model_comparison}. The MAD-based model achieved its optimum at $K = 9$ and $q = 8$, with a log-likelihood of $-504{,}975.3$ and a BIC of $1{,}031{,}607$. In contrast, the WGCNA-based model selected a smaller model with $K = 4$ and $q = 6$, yielding a substantially lower log-likelihood ($-630{,}297.4$) and a much higher BIC ($1{,}280{,}292$). The large difference in BIC ($\Delta \mathrm{BIC} \approx 250{,}000$) indicates a markedly poorer fit for the WGCNA-based model, suggesting that the reduced number of modules leads to a significant loss of information in the representation.

\begin{table}[htbp]
\centering
\caption{Model comparison between MAD-HC and WGCNA-based module construction.}
\label{tab:model_comparison}
\begin{tabular}{lcccccc}
\hline
Model & $K$ & $q$ & log-likelihood & BIC & Modules & Genes \\
\hline
MAD-based model & 9 & 8 & $-504{,}975.3$ & $1{,}031{,}607$ & 19 & 979 \\
WGCNA-based model & 4 & 6 & $-630{,}297.4$ & $1{,}280{,}292$ & 3 & 975 \\
\hline
\end{tabular}
\end{table}

Further insight is provided by the cluster composition under the WGCNA-based model, as shown in Table~\ref{tab:cluster_distribution}. The four clusters contain mixed proportions of RA and SLE samples. For example, cluster 3 includes 223 RA and 137 SLE samples, while cluster 4 contains 151 RA and 125 SLE samples. Even the smaller clusters (clusters 1 and 2) are dominated by RA but still include a non-negligible number of SLE samples. Overall, no cluster exhibits clear disease-specific separation, indicating that the WGCNA-derived modules fail to capture meaningful disease heterogeneity.

\begin{table}[htbp]
\centering
\caption{Distribution of RA and SLE samples across clusters under the WGCNA-based model.}
\label{tab:cluster_distribution}
\begin{tabular}{lcccc}
\hline
Disease & Cluster 1 & Cluster 2 & Cluster 3 & Cluster 4 \\
\hline
RA  & 73 & 46 & 223 & 151 \\
SLE & 25 & 5  & 137 & 125 \\
\hline
\end{tabular}
\end{table}

Taken together, these results suggest that the coarse module structure produced by WGCNA limits the expressive capacity of the downstream model. With only three modules, the effective dimensionality of the data is substantially reduced, constraining the model to identify fewer clusters and resulting in inferior model fit. In contrast, the finer-grained module structure obtained via MAD-HC preserves more detailed variation, enabling higher-resolution clustering and significantly improved performance.

\section{Discussion}\label{discussion6}

High-dimensional transcriptomic data present substantial challenges for unsupervised subtype discovery due to the coexistence of extreme dimensionality, complex dependence structures, and pronounced biological heterogeneity~\citep{aghaieabiane2024sgcp}. In this study, we developed a module-structured mixture factor model that incorporates gene-module information directly into both the mean and covariance components of a finite mixture framework. By modeling cluster-specific variation at the module level while accommodating latent dependence among genes, the proposed approach provides an interpretable and computationally tractable framework for analyzing heterogeneous transcriptomic populations.

A key feature of the proposed model is the explicit use of gene modules as the primary units of inference. Rather than modeling thousands of genes independently, the framework captures disease-associated variation through coordinated shifts across biologically related groups of genes. The module-structured loading formulation substantially reduces the number of covariance-related parameters relative to conventional factor-analytic mixture models, thereby improving estimation stability and scalability in high-dimensional settings. At the same time, the resulting module-level effects facilitate direct biological interpretation of the inferred clusters.\\

Application of the proposed framework to a large transcriptomic dataset comprising patients with RA and SLE demonstrated its ability to recover biologically meaningful structure in a fully unsupervised manner. Although disease labels were not used during model fitting, the inferred clusters showed strong correspondence with disease status, indicating that major disease-associated molecular signals emerge naturally from the data. More importantly, multiple clusters were identified within each disease, providing evidence of substantial molecular heterogeneity beyond conventional disease classifications~\citep{hubbard2023analysis,karmakar2024molecular}.

Functional enrichment analyses further revealed that the inferred clusters were characterised by distinct module-level molecular programs. RA-associated clusters were generally characterized by increased activity in modules related to biosynthetic and regulatory processes, including ribosome biogenesis, RNA processing, protein modification, and intracellular transport~\citep{holers2024distinct,jonsson2024synovial}. In contrast, SLE-associated clusters exhibited widespread suppression of modules enriched for immune signaling and inflammatory pathways~\citep{wang2024transcriptomic}, including myeloid activation, NF-$\kappa$B signaling, and T-cell receptor signaling. These findings suggest that disease heterogeneity is driven not only by differences between diseases but also by subtype-specific patterns of module activity within diseases.

The HPO enrichment analysis provided an additional layer of clinical interpretation. In particular, several modules were linked to immune- and haematological-related phenotypes, including autoimmune thrombocytopenia, lymphadenopathy, abnormal T-cell physiology, and platelet abnormalities. These phenotype associations support the biological relevance of the inferred module structure and suggest that the identified molecular subtypes may reflect clinically meaningful variation in disease manifestation.

Future work will focus on jointly learning module structure and clustering assignments within a unified modeling framework, as well as developing more flexible covariance representations that allow cluster-specific latent dependence~\citep{tommasini2023multiwgcna}. In addition, the proposed methodology can be naturally extended to other high-dimensional molecular data types, including proteomic, epigenomic, and multi-omics studies~\citep{boyd2025amend}. More broadly, the proposed framework illustrates how biologically informed structural constraints can improve the interpretability and stability of model-based clustering methods for complex molecular data.

\section*{Author Contributions}

Jinran Wu: Conceptualization, Methodology, Software, Formal analysis, Data curation, Validation, Visualization, Writing – original draft. Geoffrey J. McLachlan: Conceptualization, Methodology, Supervision, Project administration, Writing – review \& editing. Saumyadipta Pyne: Conceptualization, Biological interpretation, Supervision, Writing – review \& editing.

\section*{Acknowledgments}

This work was supported by the Australian Research Council [DP230101671].

\section*{Financial Disclosure}

None reported.

\section*{Conflicts of Interest}

The authors declare no conflicts of interest.

\makeatletter
\def\bibsection{\section*{\refname}}
\makeatother

\bibliographystyle{unsrtnat}
\bibliography{wileyNJD-Chicago}

@article{tibshirani1999clustering,
  title={Clustering methods for the analysis of {DNA} microarray data},
  author={Tibshirani, Robert and Hastie, Trevor and Eisen, Mike and Ross, Doug and Botstein, David and Brown, Pat and others},
  journal={Dept. Statist., Stanford Univ., Stanford, CA, Tech. Rep},
  year={1999}
}

@article{segal2004module,
  title={A module map showing conditional activity of expression modules in cancer},
  author={Segal, Eran and Friedman, Nir and Koller, Daphne and Regev, Aviv},
  journal={Nature Genetics},
  volume={36},
  number={10},
  pages={1090--1098},
  year={2004},
  publisher={Nature Publishing Group US New York}
}

@book{mclachlan2008algorithm,
  title={The EM algorithm and extensions},
  author={McLachlan, Geoffrey J and Krishnan, Thriyambakam},
  year={2008},
  publisher={John Wiley \& Sons}
}

@article{Li2025,
  title={Identifying functional subtypes and common mechanisms of rheumatoid arthritis and systemic lupus erythematosus},
  author={Li, Jiajun and Tang, Hao and Shang, Zhenwei and Chen, Rui and Meng, Xin and Cheng, Xiangshu and Song, Zerun and Li, Shuai and Zhang, Ruijie and Lv, Wenhua},
  journal={Genes \& Diseases},
  volume={12},
  number={5},
  pages={101527},
  year={2025},
  publisher={Elsevier}
}

@article{Cheng2024,
  title={The molecular subtypes of autoimmune diseases},
  author={Cheng, Xiangshu and Meng, Xin and Chen, Rui and Song, Zerun and Li, Shuai and Wei, Siyu and Lv, Hongchao and Zhang, Shuhao and Tang, Hao and Jiang, Yongshuai and others},
  journal={Computational and Structural Biotechnology Journal},
  volume={23},
  pages={1348--1363},
  year={2024},
  publisher={Elsevier}
}

@article{Cojocaru2011,
  title={Manifestations of systemic lupus erythematosus},
  author={Cojocaru, Manole and Cojocaru, Inimioara Mihaela and Silosi, Isabela and Vrabie, Camelia Doina},
  journal={Maedica},
  volume={6},
  number={4},
  pages={330},
  year={2011}
}

@article{Sharif2018,
  title={Rheumatoid arthritis in review: {C}linical, anatomical, cellular and molecular points of view},
  author={Sharif, Kassem and Sharif, Alaa and Jumah, Fareed and Oskouian, Rod and Tubbs, R Shane},
  journal={Clinical Anatomy},
  volume={31},
  number={2},
  pages={216--223},
  year={2018},
  publisher={Wiley Online Library}
}

@article{Martorell-Marugan2021,
  title={A comprehensive database for integrated analysis of omics data in autoimmune diseases},
  author={Martorell-Marug{\'a}n, Jordi and L{\'o}pez-Dom{\'\i}nguez, Ra{\'u}l and Garc{\'\i}a-Moreno, Adri{\'a}n and Toro-Dom{\'\i}nguez, Daniel and Villatoro-Garc{\'\i}a, Juan Antonio and Barturen, Guillermo and Mart{\'\i}n-G{\'o}mez, Adoraci{\'o}n and Troule, Kevin and G{\'o}mez-L{\'o}pez, Gonzalo and Al-Shahrour, F{\'a}tima and others},
  journal={BMC Bioinformatics},
  volume={22},
  number={1},
  pages={343},
  year={2021},
  publisher={Springer}
}

@article{Shen2022,
  title={A transcriptome atlas and interactive analysis platform for autoimmune disease},
  author={Shen, Zhuoqiao and Fang, Minghao and Sun, Wujianan and Tang, Meifang and Liu, Nianping and Zhu, Lin and Liu, Qian and Li, Bin and Sun, Ruoming and Shi, Yu and others},
  journal={Database},
  volume={2022},
  pages={baac050},
  year={2022},
  publisher={Oxford University Press UK}
}

@article{Langfelder2008,
  title={{WGCNA: An R} package for weighted correlation network analysis},
  author={Langfelder, Peter and Horvath, Steve},
  journal={BMC Bioinformatics},
  volume={9},
  number={1},
  pages={559},
  year={2008},
  publisher={Springer}
}

@article{Kohler2021,
  title={The human phenotype ontology in 2021},
  author={K{\"o}hler, Sebastian and Gargano, Michael and Matentzoglu, Nicolas and Carmody, Leigh C and Lewis-Smith, David and Vasilevsky, Nicole A and Danis, Daniel and Balagura, Ganna and Baynam, Gareth and Brower, Amy M and others},
  journal={Nucleic Acids Research},
  volume={49},
  number={D1},
  pages={D1207--D1217},
  year={2021},
  publisher={Oxford University Press}
}

@article{Patino-Martinez2026,
  title={Longitudinal multiomic and spatial transcriptomic profiling of lupus nephritis progression in a murine model},
  author={Patino-Martinez, Eduardo and Hajihosseini, Morteza and Hanata, Norio and Jiang, Kan and Oguz, Cihan and Tandon, Mayank and Schaughency, Paul and Randazzo, Davide and Naz, Faiza and Dell’Orso, Stefania and others},
  journal={The Journal of Immunology},
  volume={215},
  number={5},
  pages={vkag100},
  year={2026},
  publisher={Oxford University Press}
}

@article{Mi2019,
  title={Disease classification via gene network integrating modules and pathways},
  author={Mi, Zhilong and Guo, Binghui and Yin, Ziqiao and Li, Jiahui and Zheng, Zhiming},
  journal={Royal Society Open Science},
  volume={6},
  number={7},
  pages={190214},
  year={2019},
  publisher={The Royal Society}
}

@article{mclachlan2019finite,
  title={Finite mixture models},
  author={McLachlan, Geoffrey J and Lee, Sharon X and Rathnayake, Suren I},
  journal={Annual Review of Statistics and its Application},
  volume={6},
  number={1},
  pages={355--378},
  year={2019},
  publisher={Annual Reviews}
}

@book{mclachlan2005analyzing,
  author    = {McLachlan, Geoffrey J. and Do, Kim-Anh and Ambroise, Christophe},
  title     = {Analyzing Microarray Gene Expression Data},
  year      = {2004},
  publisher = {John Wiley \& Sons},
  address   = {Hoboken, NJ},
  series    = {Wiley Series in Probability and Statistics}
}

@article{mclachlan2003modelling,
  title={Modelling high-dimensional data by mixtures of factor analyzers},
  author={McLachlan, Geoffrey J and Peel, David and Bean, Richard W},
  journal={Computational Statistics \& Data Analysis},
  volume={41},
  number={3-4},
  pages={379--388},
  year={2003},
  publisher={Elsevier}
}

@inproceedings{mclachlan2000mixtures,
  title={Mixtures of factor analyzers},
  author={McLachlan, Geoffrey J and Peel, David},
  booktitle={Proceedings of the Seventeenth International Conference on Machine Learning},
  pages={599--606},
  year={2000}
}

@article{baek2009mixtures,
  title={Mixtures of factor analyzers with common factor loadings: {A}pplications to the clustering and visualization of high-dimensional data},
  author={Baek, Jangsun and McLachlan, Geoffrey J and Flack, Lloyd K},
  journal={IEEE Transactions on Pattern Analysis and Machine Intelligence},
  volume={32},
  number={7},
  pages={1298--1309},
  year={2009},
  publisher={IEEE}
}

@article{mclachlan2011mixtures,
  title={Mixtures of Factor Analysers for the Analysis of High-Dimensional Data},
  author={McLachlan, Geoffrey J and Baek, Jangsun and Rathnayake, Suren I},
  journal={Mixtures: Estimation and Applications},
  pages={189--212},
  year={2011},
  publisher={Wiley Online Library}
}

@incollection{ng2011algorithm,
  title={The {EM} algorithm},
  author={Ng, Shu Kay and Krishnan, Thriyambakam and McLachlan, Geoffrey J},
  booktitle={Handbook of Computational Statistics: {C}oncepts and Methods},
  pages={139--172},
  year={2011},
  publisher={Springer}
}

@article{aghaieabiane2024sgcp,
  title={{SGCP: A} spectral self-learning method for clustering genes in co-expression networks},
  author={Aghaieabiane, Niloofar and Koutis, Ioannis},
  journal={BMC Bioinformatics},
  volume={25},
  number={1},
  pages={230},
  year={2024},
  publisher={Springer}
}

@article{hubbard2023analysis,
  title={Analysis of transcriptomic features reveals molecular endotypes of {SLE} with clinical implications},
  author={Hubbard, Erika L and Bachali, Prathyusha and Kingsmore, Kathryn M and He, Yisha and Catalina, Michelle D and Grammer, Amrie C and Lipsky, Peter E},
  journal={Genome Medicine},
  volume={15},
  number={1},
  pages={84},
  year={2023},
  publisher={Springer}
}

@article{karmakar2024molecular,
  title={Molecular profiling and therapeutic tailoring to address disease heterogeneity in systemic lupus erythematosus},
  author={Karmakar, Abhibroto and Kumar, Uma and Prabhu, Smitha and Ravindran, Vinod and Nagaraju, Shankar Prasad and Suryakanth, Varashree Bolar and Prabhu, Mukhyaprana M and Karmakar, Subhradip},
  journal={Clinical and Experimental Medicine},
  volume={24},
  number={1},
  pages={223},
  year={2024},
  publisher={Springer}
}

@article{holers2024distinct,
  title={Distinct mucosal endotypes as initiators and drivers of rheumatoid arthritis},
  author={Holers, V Michael and Demoruelle, Kristen M and Buckner, Jane H and James, Eddie A and Firestein, Gary S and Robinson, William H and Steere, Allen C and Zhang, Fan and Norris, Jill M and Kuhn, Kristine A and others},
  journal={Nature Reviews Rheumatology},
  volume={20},
  number={10},
  pages={601--613},
  year={2024},
  publisher={Nature Publishing Group UK London}
}

@article{wang2024transcriptomic,
  title={Transcriptomic studies unravel the molecular and cellular complexity of systemic lupus erythematosus: {A} review},
  author={Wang, Frank Qingyun and Dang, Xiao and Yang, Wanling},
  journal={Clinical Immunology},
  volume={268},
  pages={110367},
  year={2024},
  publisher={Elsevier}
}

@article{tommasini2023multiwgcna,
  title={{multiWGCNA: An R} package for deep mining gene co-expression networks in multi-trait expression data},
  author={Tommasini, Dario and Fogel, Brent L},
  journal={BMC Bioinformatics},
  volume={24},
  number={1},
  pages={115},
  year={2023},
  publisher={Springer}
}

@article{wang2025stmodule,
  title={{STModule: I}dentifying tissue modules to uncover spatial components and characteristics of transcriptomic landscapes},
  author={Wang, Ran and Qian, Yan and Guo, Xiaojing and Song, Fangda and Xiong, Zhiqiang and Cai, Shirong and Bian, Xiuwu and Wong, Man Hon and Cao, Qin and Cheng, Lixin and others},
  journal={Genome Medicine},
  volume={17},
  number={1},
  pages={18},
  year={2025},
  publisher={Springer}
}

@article{silkwood2024leveraging,
  title={Leveraging gene correlations in single cell transcriptomic data},
  author={Silkwood, Kai and Dollinger, Emmanuel and Gervin, Joshua and Atwood, Scott and Nie, Qing and Lander, Arthur D},
  journal={BMC Bioinformatics},
  volume={25},
  number={1},
  pages={305},
  year={2024},
  publisher={Springer}
}

@article{jonsson2024synovial,
  title={Synovial tissue insights into heterogeneity of rheumatoid arthritis},
  author={Jonsson, Anna Helena},
  journal={Current Rheumatology Reports},
  volume={26},
  number={3},
  pages={81--88},
  year={2024},
  publisher={Springer}
}

@article{boyd2025amend,
  title={{AMEND 2.0: M}odule identification and multi-omic data integration with multiplex-heterogeneous graphs},
  author={Boyd, Samuel S and Slawson, Chad and Thompson, Jeffrey A},
  journal={BMC Bioinformatics},
  volume={26},
  number={1},
  pages={39},
  year={2025},
  publisher={Springer}
}

@article{meng1993maximum,
  title={Maximum likelihood estimation via the {ECM} algorithm: {A} general framework},
  author={Meng, Xiao-Li and Rubin, Donald B},
  journal={Biometrika},
  volume={80},
  number={2},
  pages={267--278},
  year={1993},
  publisher={Oxford University Press}
}

\section*{Supporting Information}

All data and code required to reproduce the analyses presented in this paper are publicly available at:
\url{https://github.com/wujrtudou/structured_MFA.git}.

\appendix

\section{Module structure construction and parameter settings}
\label{app:module_settings}

Gene modules were constructed according to the procedure described in Section~\ref{sec:modules}. For the autoimmune disease dataset, the top 3000 genes ranked by MAD were retained. Hierarchical agglomerative clustering with average linkage and dissimilarity measure $d_{jj'}$ was then applied. The dendrogram was cut at $r_0=0.3$, and modules containing fewer than 20 genes were removed. This resulted in 979 genes grouped into 19 modules. Model selection was performed using BIC over a grid of candidate values for the number of clusters $K$ and latent factor dimension $q$. The selected model had $K=9$ and $q=8$.
\end{document}